# On Reliability of Dynamic Addressing Routing Protocols in Mobile Ad Hoc Networks

Marcello Caleffi, Giancarlo Ferraiuolo, Luigi Paura
Department of Electronic and Telecommunication Engineering (DIET)
University of Naples "Federico II"
via Claudio 21, Naples, 80125, Italy
Tel +39 (0)81 7683810 – Fax +39 (0)81 7683149
{name.surname}@unina.it

*Abstract*- **In this paper, a *reliability* analysis is carried out to state a performance comparison between two recently proposed proactive routing algorithms. These protocols are able to scale in ad hoc and sensor networks by resorting to dynamic addressing, to face with the topology variability, which is typical of ad hoc, and sensor networks. Numerical simulations are also carried out to corroborate the results of the analysis.**

I. INTRODUCTION

In Mobile Ad hoc NETworks (MANETs) and sensor networks the scalability is a critical requirement if these technologies have to reach their full potential. However, most of experimented routing protocols have shown to work satisfactorily only up to few hundred nodes [1]. In fact, such protocols, based on traditional routing procedures, assume that node identity equals node routing address exploiting so static addressing schemes, regardless their belonging class (proactive, reactive or the hybrid one). Such an assumption is certainly unacceptable for ad hoc and sensor networks, due to the node mobility and/or link instability. The need of tracking each node position (*location management* problem) gives rise to a massive overhead problem as the network significantly grows. Recently, several works have suggested to separate the time-invariant node identity from the routing address, which is transient and reflects the node topological position inside the network. Since this approach, referred to as *dynamic addressing*, needs a mechanism to provide a scalable mapping between node identity and routing address, Distributed Hash Tables (DHTs) have been utilized. More specifically, in [2-3] a logical tree structure, based on the address space and built on connectivity among the nodes, is introduced. Although this structure allows one to perform a simple and manageable routing procedure, it lacks for robustness against mobility and/or link failures and, moreover, exhibits unsatisfactory route selection flexibility [5].

Very recently, in order to both maintain a light mechanism for address allocation and to face with the above mentioned lack of complete topological information, a routing protocol, referred to as the Augmented Tree-based Routing (ATR) protocol, has been proposed [6]. The ATR protocol augments the tree-based address allocation scheme of DART protocol [4], by storing in the node routing tables additional information, i.e. multiple routes towards the same subset of nodes. Each node acquires this information simply by using the underlying neighbor discovering procedure, without increasing therefore the protocol overhead, with respect to DART one, and with limited costs in terms of memory requirements on the node. The advantage of this approach is that a richer network-topology knowledge can be exploited to implement *temporal* multi-path strategies, which guarantee better performance and a higher *reliability* [7]. However, with a little effort, ATR could be easily extended to split data transfers on multiple paths in the *spatial* domain, in order to reduce congestion and end-to-end delay. This paper presents a reliability assessment analysis to substantiate the effectiveness of the multi-path approach of the ATR protocol and its superiority with respect to the shortest-path one of DART.

The outline of the paper is the following: in Section II, we shortly present the ATR and DART protocols under analysis. In Section III a framework for carrying out the reliability analysis is introduced, whereas in Section IV numerical performance analysis and comparison, using the framework presented in Section III and via simulations, are presented. Finally, in Section V conclusions are drawn.

II. DYNAMIC ADDRESSING ROUTING PROTOCOLS

We give here only essential information on DART and ATR protocols, and remind to [4,6] for details. In few words, ATR and DART are based on the same address space structure which can be represented by a binary tree of $l+1$ levels, where $l$ is the number of bits used for an address (Fig. 1)

ATR and DART protocols differ from the packet forwarding process and from the way in which the node routing tables are populated. More specifically, in DART each node maintains only one possible *next hop* toward the final destination, defining so a unique route along the tree structure of the address space, whereas in ATR each node maintains and explores all the possible ways to reach the final destination, through its neighbors. This is equivalent to use an *augmented tree* structure to perform forwarding, obtained without additional overhead with respect to DART. The goal

This work is partially supported both by National project Wireless 8O2.16 Multi-antenna mEsh Networks (WOMEN) under grant number 2005093248 and by the Italian Ministry of University (MIUR) project S.Co.P.E..

of this paper is to state the effectiveness of the multi-path approach, with respect to classic shortest-path one, by a theoretical reliability analysis.

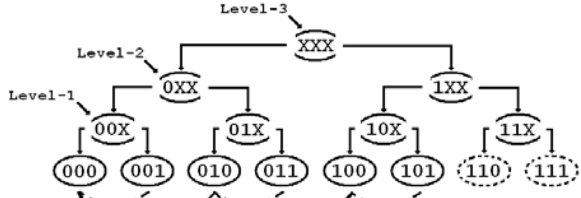

**Figure 1 - Address space structure**

### III. THEORETICAL RELIABILITY ANALYSIS

#### A. Definitions and assumptions

With reference to unicast routing, let us define the *terminal pair routing reliability* as the probability that at least one route between a couple of nodes exists:

$$R_{st}(G_P) = P(\text{nodes s and t are connected}) \quad (1)$$

where $s$ is the source node, $t$ is the destination one and $G_p=(V,E)$ is the probabilistic direct graph that represents the network topology, in which a vertex $v_i \in V$ denotes a node belonging to the network, an edge $e_{ij} \in E$ denotes the communication link between nodes $v_i$ and $v_j$ that operates with probability $p_{ij}$. The failure events of the edges $e_{ij}$ are assumed to be statistically independent of each other with probability $q_{ij}=1-p_{ij}$, while the vertexes are considered to be flawless, i.e. operative with probability one [8].

Although in real networks, there may also be outage due to finite capacity effects caused by physical-layer and link-layer constraints, as well as routing related ones, network reliability assumes that there is no routing or capacity constraint on the network. If at least one route exists topologically between $s$ and $t$, than it is assumed the packets will discover and use it. Moreover, we suppose that the network topology is static, namely the packet delivery time intervals are smaller than the topology variation ones [12]. With such an assumption, we use reliability measure as valuable tool to analyze the tolerance of routing protocols against the route failures.

#### B. Exact algorithm for routing reliability

To evaluate the network reliability from a routing point of view, we should distinguish between *physical* and *overlay* graphs related to a network. Let's start with an example. In Fig. 2 we have represented the adjacency matrixes associated with the physical and overlay graphs referring to the same network with 8 nodes. These matrixes differ only from the numbers of '1' (communication links). The matrix on the left refers to the physical graph, i.e. the graph in which the edge $e_{ij}$ is present if a physical communication link is present between the nodes $i$ and $j$. The other two matrixes represent the overlay graphs built upon the physical network by DART and ATR route discovery processes. The absence of ten edges ('1') in the DART matrix, with respect to the physical and ATR ones, evidences the inability of shortest-path routing protocols to build a complete topological view of the network

```
0 1 0 0 0 0 1 0      0 1 0 0 0 0 1 0      0 1 0 0 0 0 1 0
1 0 0 0 1 1 1 1      1 0 0 0 1 1 0 1      1 0 0 0 1 1 1 1
0 0 0 1 1 0 1 0      0 0 0 1 1 0 1 0      0 0 0 1 1 0 1 0
0 0 1 0 1 0 1 0      0 0 1 0 1 0 1 0      0 0 1 0 1 0 1 0
0 1 1 1 0 1 1 1      0 1 0 0 0 0 1 0      0 1 1 1 0 1 1 1
0 1 0 0 1 0 1 1      0 1 0 0 1 0 1 0      0 1 0 0 1 0 1 1
1 1 1 1 1 1 0 1      1 1 1 0 0 0 0 0      1 1 1 1 1 1 0 1
0 1 0 0 1 1 1 0      0 1 0 0 1 1 1 0      0 1 0 0 1 1 1 0
```
   Physical            DART logical        ATR logical
adyacency matrix   adyacency matrix  adyacency matrix

**Figure 2 - Adjacency matrix for a 8 nodes network.**

The edge set $E$ of an overlay graph could be defined as:

$$E = \bigcup_{s,t \in V} E_{st} \quad (2)$$

where $E_{st} = \{e_{ij} \in P_{st}\}$ and $P_{st}$ is the collection of *s-t* paths discovered by the routing protocol.

Fig. 3 shows the overlay graphs associated with the different route discovery results for the same full-mesh 4 nodes network. More specifically, the graphs show the routes from each node towards two destinations, say node '2' and '4'. This kind of representation underlines some notable aspects of the route discovery process. The first is the presence (or not) of multiple paths towards the same destination. Further, it allows one to recognize if the multiple paths are disjoint or partially disjoint and how many links are shared by the routes. Finally, it shows that a hierarchical shortest-path routing protocol as DART one does not find every time the shortest route, due to its hierarchical nature (in the example, the shortest route is the single hop one).

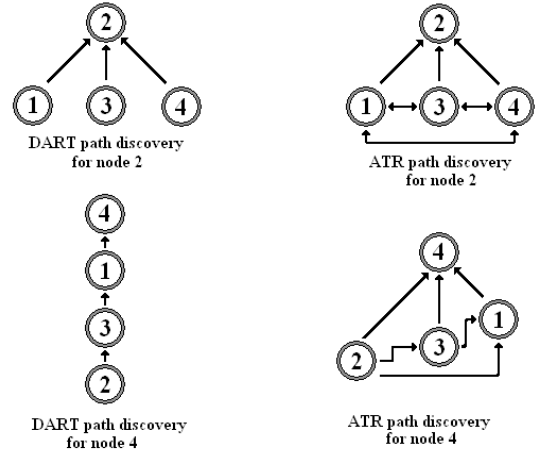

**Figure 3 – Graphs referring to route discovery process.**

Our approach to evaluate the routing reliability is based on enumerating all the minimal edge cut sets of the graph representing the network, where a minimal cut set is defined as a minimal set of elements whose failure implies that some nodes cannot communicate. We have generalized the algorithm presented in [9] in order to obtain a symbolic expression for the routing reliability, which is function of the link failure probability. This allows one to easy evaluate the reliability in different environmental conditions.

Analytically, the mean terminal-pair routing reliability is:

$$R = \frac{\sum_s \sum_{t \neq s} z_{st} R_{st}}{n(n-1)} \quad (3)$$

where $n = |V|$, $z_{st}$ is the probability of a data flow between $\{s,t\}$ and $R_{st}$ is the terminal pair routing reliability defined as:

$$R_{st}(G,p) = 1 - \sum_{i=c}^{m} C_i(G,s,t) p^{m-i}(1-p)^i \quad (4)$$

where $m = |E|$, $G=(V,E_{st})$ is the overlay graph generated by the routing discovery process, $p \equiv p_{ij}$ is the link success probability (assumed for simplicity the same for any pair of nodes), $c$ is the minimum cut of the graph between $\{s,t\}$ and $C_i$ is the number of $\{s,t\}$ cut sets composed exactly by $i$ edges.

Listing 1 shows our proposal to exact compute both symbolic and numerical reliability (4). Further detail can be found in [9].

**Listing 1 – Recursive(G,HASH,SS,n,t,notRel,symb)**

```
// Reliability = 1 – Recursive(…) output
// G is the adjacency matrix related to the overlay graph
// HASH is a collection of minimal cut set initialized to empty
// SS is the under analysis minimal cut set initialized to empty
// n is initialized to s
if (n == t) return;
merge(G, SS, n);      // Merging node n in SS
absorb(G, SS, t);     // Absorbing redundant nodes in SS
if (HASH.isPresent(SS)) return;
HASH.insert(SSt);
find a cutset C of SS;
symbTemp = "(1-p)^" + C.size.toString;
temp = 1.0;
symbTemp = symb + " + p * (" + symbTemp;
for each edge in C
    temp = pFailed * temp;
end
for each node adjacent to SS
    Recursive(G,HASH,SS,n,t,temp,symb);
    temp = pSuccess * temp;
end
symbTemp = symbTemp + ")";
notRel = notRel + temp;
```

### C. Considerations on computational times

The problem of calculating the routing reliability as terminal-pair reliability is computationally hard, even in special cases, since it belongs to the class of *#P−complete* problems [10], which are at least as difficult as the ones in the class of *NP-complete* problems. Therefore, we are looking for some useful bounds, which make easy to analyze the gain of a multi-path approach with respect to the traditional shortest-path one. Work is in progress to prove that the reliability of any shortest-path routing protocol could be upper bound by a power of $p$, where the exponent is related to the number of nodes in the network. Moreover, we are trying to bound the reliability for a multi-path routing protocol with a polynomial expression.

## IV. ROUTING PERFORMANCE ANALYSIS

In this section, we analyze the performance of DART and ATR protocols by assessing the reliability (3) and we verify that the results of such theoretical analysis agree with the ones based on a traditional metric such as the packet delivery ratio obtained via numerical simulations.

### A. Exact algorithm for routing reliability

The Listing 1 takes as input the adjacency matrix associated with the overlay graph, which is, as illustrated in Section III, the overlay graph is the result of the route discovery process performed by the routing protocol. In order to generate the overlay graph, we have tested DART and ATR via ns-2 network simulator [11] (see Section IV.B for further details) with static topologies. Then we have extracted from each node the paths information embedded in the routing table. This information has been exploited to build the overlay graphs utilized to compute the mean network reliability as in (3) and its standard deviation as function of p.

Fig. 4 compares the mean network reliability (3) of DART and ATR for a full-mesh topology with four nodes. The results confirm the capabilities of ATR multi-path approach to take advantage of redundant routes also in presence of a few nodes. The effectiveness of the multi-path approach is particularly marked for values of *p* near '0.5'.

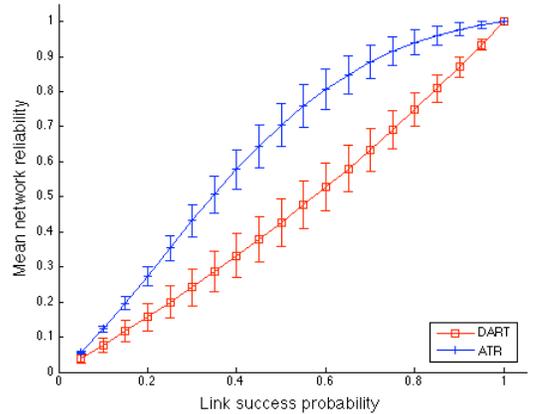

**Figure 4 - Reliability for 4 nodes full mesh topology**

In Fig. 5 we present the reliability (3) for a random topology with 16 nodes and a density of 64 nodes/Km$^2$. This figure shows that the ATR significantly outperforms DART, thanks to the availability of multiple paths. Moreover, we observe that the reliability of ATR for the 16 nodes topology (Fig. 5) is higher than that obtained for a topology with 4 nodes (Fig. 4) for any value of *p*. This fact clearly evidences that the multi-path approach becomes more advantageous when the number of redundant paths scales.

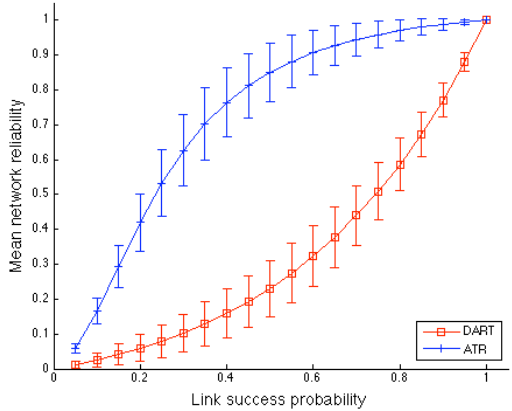

**Figure 5 – Reliability for high density 16 nodes topology**

Fig. 6 illustrates the reliability (3) for a random topology with 64 nodes and a density of 25 nodes/Km$^2$. In this case, a lower number of redundant paths, compared with the 16 nodes topology (Fig. 5), gives rise to lower values of ATR reliability. Similarly, the DART performance is lower then that associated with respect to the higher node density case (Fig. 5).

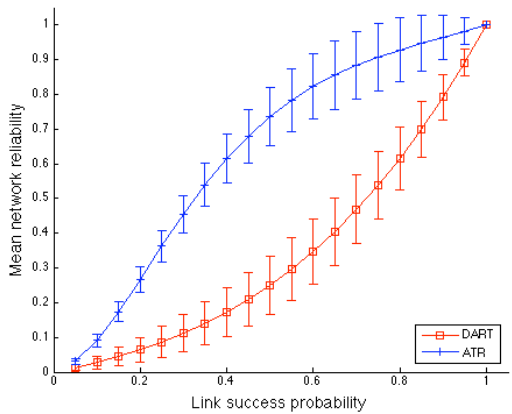

**Figure 6 - Reliability for low density 64 nodes topology**

*B. Simulation results*

In this sub-section, we present results of a performance analysis of the proposed routing protocols, achieved by numerical experiments carried out by resorting to *ns-2* (version 2.30) network simulator [11]. We adopt the standard values for both the physical and the link layer to simulate an IEEE 802.11a Lucent network interface with *Two-Ray Ground* as channel model (where the link success probability is '1' in the transmission range).

Here, we present only simulation results to verify that the results of the theoretical reliability analysis previously reported agree with the ones based on traditional routing metric, such as packet delivery ratio. A more detailed performance comparison of the two protocols can be found in [6].

For all the simulations, we adopt the *Random Waypoint* mobility model, with the speed values randomly taken in the [0.5m/s; 5m/s] range, and the pause time in [0s; 100s]. Although much more realistic models are available in literature, we have adopted the Random Waypoint one since, due to its simplicity, it has become a standard choice. Each trial is 750 seconds long, with the first 450 free of mobility and data traffic, dedicated to the address allocation process. The size of simulation area is chosen in order to keep steady the node density. Simulation results refer to a density of 64 nodes/km$^2$, which corresponds to a node connectivity degree of 12; this value guarantees for most of the cases a connected topology. The data traffic is modeled as CBR flows over UDP protocol and the global throughput is kept constant at 250Kbit/s. The start and the end time of each flow are randomly selected in the interval [450s; 730s] according to a uniform distribution, and the number of flows grows with the number of nodes *N*. Since we do not deal with DHT layer, this is replaced in the simulations by a global known table.

Here we compare DART and ATR using the packet delivery ratio. Fig. 7 shows the ratio of correctly delivered packets with respect to the total number of sent packets for both DART and ATR, normalized to the ATR values, versus the number of nodes. The experimental results show that ATR scales always better than DART in terms of packet delivery ratio, due to its multi-path characteristic. Let us underline that, as the number of nodes grows, the link failures, due to mobility and/or collision effects, make the multi-path approach more effective with respect to the shortest-path one, confirming so the reliability analysis results of Section 4.A.

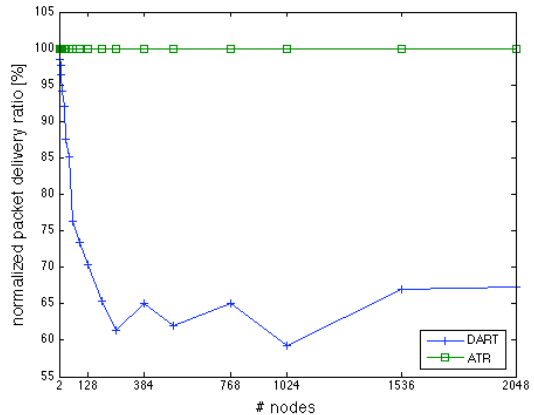

**Figure 7 - Normalized packet delivery ratio**

V. CONCLUSIONS AND FUTURE WORKS

In this paper, a theoretical reliability analysis of recently proposed DART and ATR routing protocols for MANETs and sensor networks is presented. An effective method to evaluate routing reliability is proposed, and it is used to compare the two considered protocols. The results of the theoretical analysis are also in agreement with those based on a traditional metric, such as the packet delivery ratio obtained by numerical simulations. More specifically, the results show that ATR multi-path approach is suitable for feasible routing and works always significantly better than DART shortest-path one in large networks. Let us underline that the proposed method for reliability evaluation can be considered as a more general framework to analyze routing protocols, and current work is in progress in such direction.

REFERENCES

[1] Broch, J., Maltz, D.A., Johnson, D.B., Hu, Y. and Jetcheva, J., "A performance comparison of multi-hop wireless ad hoc network routing protocols", in *MobiCom '98: Proceedings of the 4th annual ACM/IEEE international conference on Mobile computing and networking*, 1998, pp. 85-97.

[2] Gerla, M., Hong, X. and Pei, G., "Landmark routing in ad hoc networks with mobile backbones", in *Journal of Parallel and Distributed Computing*, vol. 63, no. 2, pp. 110-122, February 2003.

[3] Viana, A. C., de Amorim, M. D., Fdida, S. and de Rezende, J. F., "Indirect routing using distributed location information", in *PERCOM '03: Proceedings of the First IEEE International Conference on Pervasive Computing and Communications*, 2003, pp. 224.

[4] J. Eriksson, M. Faloutsos and S. Krishnamurthy. "DART: Dynamic Address RouTing for Scalable Ad Hoc and Mesh Networks". in *IEEE- ACM Transactions on Networking*, vol. 15, no. 1, April 2007, pp. 119-132.

[5] Alvarez-Hamelin, J.I., Viana, A.C.; De Amorim, M.D., "Architectural Considerations for a Self-Configuring Routing Scheme for Spontaneous Networks", in *Technical Report*, vol. 1, October 2005, pp. 1.

[6] Caleffi, M., Ferraiuolo, G., and Paura, L., "Augmented Tree-based Routing Protocol for Scalable Ad Hoc Networks", in *MHWMN '07: Proceedings of the Third IEEE International Workshop on Heterogeneous Multi-Hop Wireless and Mobile Networks*, 2007.

[7] Lee, S.J., and Gerla, M., "Split Multipath Routing with Maximally Disjoint Paths in Ad Hoc Networks", in *ICC '01: Proceedings of the IEEE International Conference on Communications*, pp. 3201-3205, 2001

[8] Ball., M.O., "Complexity of network reliability computations", in *Networks*, vol. 10, no. 2, 1980, pp. 153-165.

[9] Lin, H., Kuo, S., and Yeh, F., "Minimal cutset enumeration and network reliability evaluation by recursive merge and BDD", in *ISCC '03: Proceedings of the 8th IEEE international Symposium on Computers and Communications*, 2003, pp. 1341-1346.

[10] Valiant, L.G., "The complexity of enumeration and reliability problems", in *SIAM Journal of Computing*, vol. 9, 1979, pp. 410-421.

[11] The VINT Project. "The ns Manual (formerly ns Notes and Documentation)".

[12] Bai, F., Sadagopan, N., Krishnamachari, B., and Helmy, A., "Modeling path duration distributions in MANETs and their impact on reactive routing protocols", in IEEE Journal on Selected Areas in Communications, vol. 22, no. 7, 2004, pp. 1357-1372.